\documentclass[12pt, preprint]{emulateapj}
\usepackage{graphicx}
\usepackage[]{natbib}
\def \ergcmsec{\hbox{erg cm$^{-2}$ s$^{-1}$}}
\def \phcmsec{\hbox{photons cm$^{-2}$ s$^{-1}$}}

\def \gray {$\gamma$-ray }
\def \source {\hbox{3C~454.3}}
\def \agile {AGILE}

\def \swi {{\it Swift}}

\def \suzaku {{\it Suzaku}{}}
\def \MITSUME {MITSuME}
\usepackage{subfigure}
\begin{document}
\slugcomment{accepted by ApJ}
   \title{Multiwavelength observations of 3C 454.3
II. The \agile{} 2007 December campaign}

   \author{I. Donnarumma\altaffilmark{1,*},
 G. Pucella\altaffilmark{1},
V.~Vittorini\altaffilmark{1}, F.~D'Ammando\altaffilmark{1,2}, S.
Vercellone\altaffilmark{3},
  C. M.~Raiteri\altaffilmark{4}, M.~Villata\altaffilmark{4}, M.~Perri\altaffilmark{22}, W.P. Chen\altaffilmark{5}, R.L. Smart\altaffilmark{4},
  J.~Kataoka\altaffilmark{6}, N. Kawai\altaffilmark{6}, Y. Mori\altaffilmark{6},
  G. Tosti\altaffilmark{7}, D. Impiombato\altaffilmark{7},
  T. Takahashi\altaffilmark{8}, R. Sato\altaffilmark{8},
  M.~Tavani\altaffilmark{1,2}, A.~Bulgarelli\altaffilmark{9},
  A.~W.~Chen\altaffilmark{10}, A.~Giuliani\altaffilmark{10},
  F.~Longo\altaffilmark{11}, L.~Pacciani\altaffilmark{1},
  A.~Argan\altaffilmark{1}, G.~Barbiellini\altaffilmark{11},
  F.~Boffelli\altaffilmark{12}, P.~Caraveo\altaffilmark{10},
  P.~W.~Cattaneo\altaffilmark{12}, V.~Cocco\altaffilmark{1}, T.~Contessi\altaffilmark{10},
  E.~Costa\altaffilmark{1}, E.~Del Monte\altaffilmark{1}, G.~De
  Paris\altaffilmark{1}, G.~Di Cocco\altaffilmark{9},
  Y.~Evangelista\altaffilmark{1}, M.~Feroci\altaffilmark{1},
  A.~Ferrari\altaffilmark{13,14}, M.~Fiorini\altaffilmark{10}, T. Froysland\altaffilmark{13},
  M.~Frutti\altaffilmark{1},  F.~Fuschino\altaffilmark{9}, M.~Galli\altaffilmark{15},
 F.~Gianotti\altaffilmark{9},
  C.~Labanti\altaffilmark{9}, I.~Lapshov\altaffilmark{1},
 F.~Lazzarotto\altaffilmark{1},
  P.~Lipari\altaffilmark{16},
    M.~Marisaldi\altaffilmark{9}, M.~Mastropietro\altaffilmark{17},
    S.~Mereghetti\altaffilmark{10}, E.~Morelli\altaffilmark{9}, E.~Moretti\altaffilmark{11},
    A.~Morselli\altaffilmark{18}, A.~Pellizzoni\altaffilmark{19}, F.~Perotti\altaffilmark{10},
  P.~Picozza\altaffilmark{18}, M. Pilia\altaffilmark{19,20}, G.~Porrovecchio\altaffilmark{1},
  M.~Prest\altaffilmark{20},  M.~Rapisarda\altaffilmark{21},
  A.~Rappoldi\altaffilmark{12}, A.~Rubini\altaffilmark{1},
  S.~Sabatini\altaffilmark{2}, E. Scalise\altaffilmark{1},
  P.~Soffitta\altaffilmark{1}, E. Striani\altaffilmark{1,2}, M.~Trifoglio\altaffilmark{9},
  A.~Trois\altaffilmark{1}, E.~Vallazza\altaffilmark{11}, A.~Zambra\altaffilmark{10},  D.~Zanello\altaffilmark{17},
  C.~Pittori\altaffilmark{22}, P.~Santolamazza\altaffilmark{22}, F.~Verrecchia\altaffilmark{22}, P.~Giommi\altaffilmark{22}, L.~A.~Antonelli\altaffilmark{23},  S.~Colafrancesco\altaffilmark{22}, L.~Salotti\altaffilmark{24}}

\altaffiltext{1}{INAF/IASF--Roma, Via del Fosso del Cavaliere 100,
  I-00133 Roma, Italy}
\altaffiltext{2}{Dip. di Fisica, Univ. ``Tor Vergata'', Via della Ricerca Scientifica 1,
  I-00133 Roma, Italy}
\altaffiltext{3}{INAF/IASF Palermo Via Ugo La Malfa 153, 90146 Palermo, Italy}
\altaffiltext{4}{INAF/OATo, Via Osservatorio 20, I-10025 Pino Torinese, Italy}
\altaffiltext{5}{Institute of Astronomy, National Central University, Taiwan}
\altaffiltext{6}{Dep. of Physics, Tokio Institute of Technology, Tokyo, Japan}
\altaffiltext{7}{Dip. di Fisica, Univ. di Perugia, Via Pascoli, I-06123 Perugia, Italy }
\altaffiltext{8}{ISAS/JAXA}
\altaffiltext{9} {INAF/IASF--Bologna, Via Gobetti 101, I-40129 Bologna, Italy}
\altaffiltext{10}{ INAF/IASF--Milano, Via E.~Bassini 15, I-20133 Milano, Italy }
\altaffiltext{11}{ Dip. di Fisica and INFN Trieste, Via Valerio 2, I-34127 Trieste, Italy}
\altaffiltext{12} {INFN--Pavia, Via Bassi 6, I-27100 Pavia, Italy}
\altaffiltext{13}{CIFS--Torino, Viale Settimio Severo 63, I-10133 Torino, Italy}
\altaffiltext{14} {Dip. di Fisica Generale dell'Universit\'a, Via Pietro Giuria 1, I-10125 Tori
no, Italy}
\altaffiltext{15}{ENEA, Via Martiri di Monte Sole 4, I-40129 Bologna,
  Italy}
\altaffiltext{16}{INFN--Roma ``La Sapienza'', Piazzale A. Moro 2, I-00185 Roma,
  Italy}
\altaffiltext{17}{CNR, Istituto Metodologie Inorganiche e dei Plasmi, Area Ricerca
  Montelibretti (Roma), Italy}
\altaffiltext{18}{INFN--Roma ``Tor Vergata'', Via della Ricerca Scientifica 1,
  I-00133 Roma, Italy}
\altaffiltext{19}{INAF-Osservatorio Astronomico di Cagliari, località Poggio dei Pini, strada 54, I-09012 Capoterra, Italy}
\altaffiltext{20}{Dip. di Fisica, Univ. dell'Insubria, Via Valleggio 11,
  I-22100 Como, Italy}
\altaffiltext{21}{ENEA--Roma, Via E. Fermi 45, I-00044 Frascati (Roma), Italy}
\altaffiltext{22}{ASI--ASDC, Via G. Galilei, I-00044 Frascati (Roma), Italy}
\altaffiltext{23}{Osservarorio Astronomico di Roma, Monte Porzio Catone (Roma), Italy}
\altaffiltext{24}{ASI, Viale Liegi 26 , I-00198 Roma, Italy}

\altaffiltext{*} {AGILE Team Corresponding Author: I. Donnarumma, immacolata.donnarumma@iasf-roma.inaf.it}

\begin{abstract}
We report on  the second \agile{} multiwavelength campaign of the blazar 3C 454.3
  during the first half of December 2007. This  campaign involved AGILE, {\it Spitzer}, {\it Swift}, {\it
    Suzaku}, the WEBT consortium, the REM and MITSuME telescopes, offering a broad band coverage that
  allowed for a simultaneous sampling of the synchrotron and inverse
  Compton (IC) emissions. The 2-week \agile{} monitoring was accompanied
  by radio to optical monitoring by WEBT and REM and by sparse observations in mid-Infrared and soft/hard X-ray energy bands
  performed by means of Target of Opportunity observations by {\it Spitzer}, {\it Swift} and {\it Suzaku}, respectively.
The source was detected with an average flux of $\sim 250
\times 10^{-8}$ \phcmsec above 100 MeV, typical of its flaring states.
 The simultaneous optical and $\gamma$-ray monitoring allowed us to
  study the time-lag associated with the variability in the two energy
bands, resulting in
  a possible $\lesssim$ 1-day delay  of the gamma-ray emission with respect to the
  optical one.
From the simultaneous 
  optical and gamma-ray fast flare detected on December 12, we can constrain
  the delay between the gamma-ray and optical emissions within 12 hours.
Moreover, we obtain three Spectral Energy Distributions (SEDs)
with simultaneous data for 2007 December 5, 13, 15, characterized
by the widest multifrequency coverage. We found
that a model with an external Compton  on seed photons
by a standard disk and reprocessed by the Broad Line Regions does not describe
in a satisfactory way the SEDs of 2007 December 5, 13 and 15. An
additional contribution, possibly from the hot corona with $T
= 10^{6}$ K surrounding the jet, is required to account
simultaneously for the softness of the synchrotron and the hardness
of the inverse Compton emissions during those epochs.

\end{abstract}
\shorttitle{Observations of 3C 454.3  in December 2007}
\shortauthors{I.Donnarumma}
\keywords{FSRQs objects: individual 3C 454.3;  radiation mechanism:
  non-thermal; X-rays: galaxies, gamma-rays: observations; galaxies: jets}

%

\section{Introduction}
3C 454.3 is a flat spectrum radio quasar at redshift $z=0.859$.
It is one of the  brightest extragalactic radio sources with superluminal
motion hosting a radio and X-ray jet.
It has been observed in almost all the electromagnetic spectrum from radio up
to \gray energies;
the SED has the typical double-humped shape of the
blazars, the first peak occurring at mid-far infrared frequencies and the
second one at MeV--GeV energies (see Ghisellini et al. 1998).

The first peak is
commonly interpreted as synchrotron radiation from high energy electrons in a
relativistic jet, while the second component is due to electrons scattering on
soft seed photons. In the context of a simple, homogeneous scenario, the emission at the
synchrotron and IC peaks is produced by the same electrons population that can
self scatter the same synchrotron photons (Synchrotron Self Compton,
SSC).
Alternatively, the jet-electrons producing the synchrotron flux can
Compton-scatter seed photons produced outside of the jet (External Compton, EC).

3C 454.3 is a highly variable blazar source. In spring 2005,  3C 454.3 experienced a strong outburst in the optical band
reaching its historical maximum with $R=12.0 $ mag  (Villata et al. 2006). The
exceptional event triggered observations at higher energies from Chandra (Villata et al. 2006), {\it Swift}
(Giommi et al. 2006) and INTEGRAL (Pian et al. 2006).  The available data
allowed to build the spectrum of the source up to 200 keV. In particular, INTEGRAL detected  (15-18 May
2005) the source from 3 up to 200 keV in a very bright state ($\sim  5\times
10^{-10}$ erg cm$^{-2}$ s$^{-1}$), being almost a factor 2-3 higher than the
previously observed fluxes (see Tavecchio et al. 2002). Pian et al. (2006) compared the SED in 2000 with that obtained during the 2005
outburst. They were able  to describe both observed SEDs with minimal changes
in the jet power, assuming that the dissipation region (where most of the
radiation is produced) was inside the Broad Line Region (BLR) in 2000 and
outside of it  in 2005. On the other hand, Sikora et al. (2008) argued that X-rays and \gray could be produced via inverse Compton scattering of
near- and mid-IR photons emitted by the hot dust: a very moderate
energy density of the dust radiation is sufficient to provide the dominance of
the EC luminosities over the SSC ones.

In July 2007 the source woke up again in the optical band, reaching a maximum at $R=12.6$  mag (Raiteri et al. 2008b).
Such an increase in the optical activity triggered observations with \swi{} and
\agile{}.  Although still in its Science Verification Phase,  \agile{}
repointed at 3C 454.3 and detected it in high
$\gamma$-ray activity.
The average \gray flux detected by \agile{} was the highest $\gamma$-ray flux
ever detected from
this blazar, being $(280 \pm 40) \times 10^{-8}$ \phcmsec (see Fig. 3 lower panel in Vercellone et al. 2008).

Ghisellini et al. (2007) reproduced the three source states in 2000, 2005, 2007  with the model proposed in
Katarzynski $\&$ Ghisellini (2007).
The model  assumes that the relative importance of synchrotron and SSC
luminosity with respect to the EC one is controlled by the value
of the bulk Lorentz factor $\Gamma$, which is associated to the compactness of the source.

Villata et al. (2007) and Raiteri et al. (2008b) suggested an alternative
interpretation involving changes of the viewing angle of the different
emitting regions of the jet.

In both cases, a strong degeneracy of
parameters exists in both SSC and  EC models especially when the \gray data are missing. This is
the case of both SEDs in 2000, 2005 in which historical EGRET data were
used to constrain the models.

3C 454.3 exhibited outbursts several times between 2007 and 2008 (see
Vercellone et al. 2008, Tosti et al. 2008, Raiteri et al. 2008a, Raiteri et
al. 2008b, Vercellone et al. 2009)  posing stringent constraints on its
gamma-ray duty cycle. 

This strengthened the need for simultaneous
observations in different energy bands. In the case of 3C 454.3 (and other
MeV blazars) it is clear that the dominant contribution in the SED comes from IR-optical bands, where
the synchrotron peak lies and from both X-rays and \gray energy range
where the inverse Compton emission lies.

In this paper we present and discuss the result of a multiwavelength campaign
on 3C 454.3 during a period of intense $\gamma$-ray activity  occurred between
2007 December 1 and 16. In Section 2 we present the multiwavelength campaign,
in Sect. 3 - 7 we present the \agile, \suzaku{}, \swi{}, $Spitzer$, REM, WEBT and \MITSUME{}
observations and data analysis; 
in Sect.\  8 we analyse the $\gamma$-optical
correlation and present broad-band SEDs built with simultaneous data,
discussing in details how they are modelled in the framework of SSC and EC
scenarios.
Throughout this paper the photon indexes are parametrized as
  $N(E)\propto E^{-\Gamma}$ (\phcmsec keV$^{-1}$ or MeV$^{-1}$). The
  uncertainies are given at 1-$\sigma$ level, unless otherwise stated.

\section{The multiwavelength campaign}

During the period of intense $\gamma$-ray activity showed in November 2007
(Vercellone et al. 2009), \agile{} continued the pointing towards 3C 454.3 for the first half of December 2007.
The persistent high $\gamma$-ray activity of the source stimulated us to activate a new
multiwavelength campaign.

\agile{} data were collected between 2007 December 1  and 2007 December
16. \suzaku{}{} data were collected during a dedicated Target of Opportunity
(ToO) performed on December 5, whereas
the {\it Spitzer} data were collected on December 13 and 15 thanks to a
granted Director's Discretionary Time (DDT) observation.

During these two days a ToO with {\it Swift} data was activated for a
total exposure of 9 ks.

During the whole \agile{} observations the source was  monitored in
radio-to-optical bands by WEBT  (see Raiteri et al. 2008a). In addition,
observations in the NIR and optical energy bands by REM occurred between
December 1 and 8. Moreover,
optical data from \MITSUME{} telescope are available on this source until December 6.
In the following sections we report on the details of the observations and the
data analysis for each instrument.

\section{\agile{} Observation}
\agile{} (Astrorivelatore Gamma a Immagini Leggero, Tavani et al. 2008, 2009) is a mission of the Italian Space Agency (ASI) for the exploration
of $\gamma$-ray sky, operating in a low Earth orbit since 2007 April 23.
The \agile{} scientific Instrument (Prest et al. 2003,
Perotti et al. 2006, Labanti et al. 2009) is very compact and
combines four active detectors yielding simultaneous coverage in gamma-rays,
30 MeV-30 GeV and in hard X-ray energy band 18-60 keV (Feroci et al. 2007).

The \agile{} observations of 3C 454.3 were performed between 2007 December 1
and 16, for a 2-week total pointing duration. In the
first period, between December 1 and 5, the source was located $\sim 45^\circ$
off the \agile{} pointing direction. In the second period, between  December 5
and 16, after a satellite re-pointing, the source was located at
$\sim 30^{\circ}$ off-axis (variable by $\sim 1$ degree per day due to
the pointing drift) thus to increase the significance of the detection.



\agile-GRID (AGILE-Gamma Rays Imaging Detector) data were analyzed using the Standard Analysis
Pipeline. Counts, exposure, and Galactic background maps were created with a bin-size of $0.3^{\circ} \times 0.3^{\circ}$\, for photons with energy greater than 100 MeV. To reduce the particle background contamination we selected only events flagged as confirmed $\gamma$-ray events, and all events collected during the South Atlantic Anomaly were rejected.
We also reduced the $\gamma$-ray Earth albedo contamination by excluding
regions within $\sim 10^{\circ}$ from the Earth limb.

The 2-week data  have been divided in 2 sets taking into account the two
different pointings during which the source shifted from $\sim 45$  to $\sim
30$ degrees off-axis: the first set between UTC 2007-12-01 13:21 and UTC
2007-12-05 12:34; the second set between UTC 2007-12-05 12:35 and 2007-12-16
10:27 (in Fig. 1, top panel,  we also report the \gray flux of 1
day before). 
The first set required a more detailed
analysis due to uncertainty on calibration for large off-axis angles in
the Field of View (FoV).  

We ran the \agile{} Maximum Likelihood procedure (ALIKE) on each data set, in order to obtain the average flux as well as
the daily values in the $\gamma$-ray band, according to Mattox et
al. (1996). The average fluxes obtained integrating separately the two data sets
are $(280\pm 50) \times 10^{-8}$ \phcmsec ($\sqrt(TS) \sim 8$) and $(210\pm16) \times 10^{-8}$ \phcmsec ($\sqrt(TS)\sim 20$) for the first and second periods, respectively. 
The source was always detected  on the 2-week period with a daily integration
time. The 1-day binned light curve shows three enhancements of the
emission around December 4,  December 7 and December 13 (see Fig. 1 top
panel). In particular, the last two enhancements are characterized by a sharp increase of the emission
followed by a slow recovery.  
We accumulated the
spectrum over the second set of data in which the source was positioned within
30 degrees in the \agile-GRID Field of View, where the most significant energy
spectrum can be extracted, due to the higher statistical quality. The spectral
fit was performed by using only data between 100 MeV and 1 GeV (which are
better calibrated) although in Fig. 2 we report also the energy bin below 100 MeV.  It resulted in a power law with a photon index
$\Gamma = 1.78\pm 0.14$. We note that the current AGILE response
is calibrated up to 1 GeV, and that the
energy flux above 1 GeV is underestimated
by a factor of 2-3. This prevented us to discuss any possible spectral break
above  1 GeV as found  by {\it Fermi} (Abdo et al. 2009). However, we note
that the AGILE spectrum seems to be harder than the one inferred by {\it
  Fermi} below $\sim 3$ GeV. The different energy range and the
non-simultaneity of the data could explain the difference between the photon
indexes.

Super\agile{} did not detect the source during the 2-week \agile{}
pointing. A deep 3-$\sigma$ upper limit of $\sim 10$ mCrab was derived integrating all the data in which
the source was within 30 degrees in the FoV (net source exposure of 360 ks).


\section{\suzaku{} Observation}

Following the \agile{} detection of the flaring state,
3C~454.3 was observed with {\suzaku{}} (Mitsuda et al.\ 2007)
on December 2007 as a ToO,  with
a total duration of 40 ks. {\suzaku{}} carries four sets of
X-ray telescopes (Serlemitsos et al.\ 2007) each one equipped with a focal-plane
X-ray CCD camera (XIS, X-ray Imaging Spectrometer; Koyama et al.\ 2007)
that is sensitive in the energy range of 0.3$-$12\,keV, together with
a non-imaging Hard X-ray Detector (HXD;  Takahashi et al.\ 2007;
Kokubun et al.\ 2007), which covers the 10$-$600\,keV energy band
with Si PIN photo-diodes and GSO scintillation detectors.
3C~454.3 was focused on the nominal center position of the
XIS detectors.

For the XIS, we analyzed the screened data, reduced via {\suzaku{}}
software version 2.1. The reduction followed the prescriptions
described in `The \suzaku{} Data Reduction Guide' provided by the {\suzaku{}} guest observer facility at the
NASA/GSFC\footnote{http://suzaku.gsfc.nasa.gov/docs/suzaku/analysis/abc.
See also seven steps to the {\suzaku{}} data analysis at
http://www.astro.isas.jaxa.jp/suzaku/analysis}.
The screening was based on the following criteria: (1) only
ASCA-grade 0, 2, 3, 4, 6 events are accumulated, while hot and
flickering pixels were removed from the XIS image using the
\textsc{cleansis} script, (2) the time interval after the passage
through the South Atlantic Anomaly (T\_SAA\_HXD) is greater than 500\,s, (3)
the object is at least 5$^\circ$ and 20$^\circ$ above the rim of the
Earth (ELV) during night and day, respectively.
In addition, we also selected the data with a cut-off rigidity (COR)
larger than 6\,GV. After this screening, the net exposure for good time
intervals is 35.1\,ks.  The XIS events were extracted from a circular
region with a radius of $4.3'$ centred on the source peak,
whereas the background was accumulated in an annulus with inner
and outer radii of $5.0'$ and $7.0'$ pixels, respectively.
The response (RMF) and auxiliary (ARF) files are produced using the
analysis tools \textsc{xisrmfgen} and \textsc{xissimarfgen}, which are
included in the software package HEAsoft version 6.4.1.

The HXD/PIN event data (version 2.1) are processed with basically
the same screening criteria as those for the XIS, except that
ELV\,$\ge$\,5$^\circ$ through night and day,
and COR\,$\ge$\,8\,GV. The HXD/PIN instrumental background spectra were
generated from a time dependent model provided by the HXD instrument
team for each observation (see Kokubun et al.\ 2007). Both the source
and background spectra were made with identical good time intervals
(GTIs) and the exposure was corrected for a detector deadtime of 6.9$\%$.
We used the response files version
\textsc{ae\_hxd\_pinxinome\_20070914.rsp}, provided by the HXD
instrumental team. Similarly, the HXD/GSO event data (version 2.1)
were processed with a standard analysis technique described in the cited
`The \suzaku{} Data Reduction Guide'.
Despite the relatively high instrumental background of the
HXD/GSO, the source was marginally detected at 5.5 $\sigma$ level
between 80 and 120 keV. We used the response files version
\textsc{ae\_hxd\_gsoxinom\_20080129.rsp}. 
Spectral analysis was performed using the Xspec fitting package 12.3.1. and we
fitted both the soft and hard X-ray spectra with a power law with Galactic
absorption free to vary.  The XIS spectra are well fitted with a
power law with $\Gamma = 1.63$  absorbed with N$_{H}= 1.1 \times 10^{21}$
cm$^{-2}$, which infers the absorbed fluxes of $4.51^{+0.07}_{-0.03}\times 10^{-11}$ erg cm$^{-2}$
s$^{-1}$ and $3.20^{+0.04}_{-0.01}\times 10^{-11}$ erg cm$^{-2}$
s$^{-1}$ in the energy bands 0.3-10 keV and 2-10 keV, respectively.
The hard X-ray spectrum determined by HXD/PIN and GSO
seems to be  a bit flatter than those determined by the XIS only below 10 keV,
as it is shown in the residuals reported in Fig. 3 (where a model with a
single power law is assumed). We found that it is better fitted by 
a power-law photon index $\Gamma$ = 1.35$\pm$0.14, which gives  F$_{10-100
  \rm keV}= 1.37^{+0.1}_{-0.08}\times 10^{-10}$ erg cm$^{-2}$
s$^{-1}$. 
The uncertainties reported above are at 90$\%$ confidence level.

\section{\swi{}{} observations}

During our campaign \swi{}{} (Gehrels et al. 2004) performed two ToO observations of 3C 454.3: the
first on 2007 December 13, the second on 2007 December 15. Both
observations were performed using all on-board experiments: the X-ray
Telescope (XRT; Burrows et al. 2005, 0.2-10 keV), the UV and Optical Telescope
(UVOT; Roming et al. 2005, 170-600 nm) and the Burst Alert Telescope (BAT;
Barthelmy et al. 2005, 15-150 keV).
The hard X-ray flux of this source is
below the sensitivity of the BAT instrument for short exposure and therefore
the data from this instruments will not be used. We refer to Raiteri et
al. (2008a) for a detailed description of data reduction and analysis of the UVOT data.

XRT observations were carried out using the instrument in Photon Counting (PC)
readout mode (see Burrows et al. 2005 and Hill 2004, for details of the XRT
observing modes). 
The XRT data were processed with the XRTDAS software package (v.2.2.2)
developed at the ASI Science Data Center (ASDC) and distributed by HEASARC
within the HEASoft package (v. 6.4). Event files were calibrated and cleaned with
standard filtering criteria with the {\it xrtpipeline}
task using the latest calibration files available in the Swift CALDB distributed by
HEASARC. Both observations showed an average count rate $>$
0.5 counts s$^{-1}$ and therefore pile-up correction was required. We extracted
the source events from an annulous extraction region with inner, outer radii
of 3, 30
pixels. To account for the background, we also extracted events within a
circular region centred on a region free from background sources and with
radius of 80 pixels.
The ancillary response files were generated with the task
xrtmkarf. We used the latest spectral redistribution matrices (RMF, v011) in
the Calibration Database maintained by HEASARC.
The adopted energy range for spectral fitting is 0.3-10 keV, and all data are
rebinned with a minimum of 20 counts per energy bin to use the $\chi^2$
statistics. \swi{}/XRT uncertainties are given at 90$\%$ confidence level for one interesting parameter, unless otherwise stated.

Spectral analysis was performed using the Xspec fitting package 12.3.1 and we
fitted the spectra with a power law model with galactic absorption left free to
vary.
In Table 1 we summarize the best fit parameters and the derived absorbed
fluxes in the energy ranges 0.3-10 keV, 2-10 keV.
We note that the best-fit N$_{H}$ values in
Table 1 are in agreement with the value $1.34 \times 10^{21}$ cm$^{-2}$ derived by Villata et al. (2006) when analysing Chandra observations in May 2005, and adopted by Raiteri et
al. (2007) and Raiteri et al. (2008b) when fitting the X-ray
spectra acquired by XMM-{\it Newton} in 2006--2007.
In Fig. 4 we show the data and the folded models for these observations.




\section{Optical monitoring}

\subsection{WEBT Observation}
The Whole Earth Blazar Telescope (WEBT; http://www.oato.inaf.it/blazars/webt/) is an
international collaboration including tens of optical, near-IR, and radio astronomers
devoted to blazar studies.
An extensive monitoring effort on 3C 454.3 was carried out by the WEBT from 2005 to 2008,
to follow the large 2005 outburst and post-outburst phases (Villata et al. 2006, 2007;
Raiteri et al. 2007), and the new flaring phase started in mid 2007 (Raiteri
et al. 2008a, 2008b). A detailed presentation and discussion of the radio, mm, optical
and {\it SWIFT}-UVOT data collected in December 2007 can be found in Raiteri et
al. (2008a). Here we adopt their data analysis in the context of our
multifrequency study. 
\subsection{REM Observation}

The photometric optical observations were carried out
with the Rapid Eye Mount (REM, Zerbi et al. 2004), a robotic telescope located
at the ESO Cerro La Silla observatory (Chile). The REM telescope has a
Ritchey-Chretien configuration with a 60 cm f/2.2 primary and an overall f/8
focal ratio in a fast moving alt-azimuth mount providing two stable Nasmyth
focal stations. Two cameras are simultaneously used at the focus of the
telescope, by means of a dichroic filter, REMIR for the NIR (Conconi et
al. 2004) and ROSS for the optical (Tosti et al. 2004), in order to obtain
nearly simultaneous data.

The telescope REM has continuously observed 3C 454.3 between
2007 December 1 and 2007 December 8, overlapping
with the \agile{} observation period. The light curve produced by REM in the $R$-band is
shown in Fig. 1 (bottom panel, red points). 

\subsection{MitSuME Observation}
A contribution of the optical follow-up observations was given also by MITSuME
(Multicolor Imaging Telescopes for Survey and Monstrous Explosions),
composed of 3 robotic telescopes (of 50 cm diameter each) located at
the ICCR (Institute of Cosmic-Ray Research) Akeno Observatory,
Yamanashi, Japan and the OAO (Okayama Astrophysical Observatory).
Each MITSuME telescope has a Tricolor Camera, which allows
to take simultaneous images in $g'$, $R_c$ and $I_c$ bands. The
camera employs three Alta U-6 cameras (Apogee Instruments Inc.) and
KAF-1001E CCD (Kodak) with 1024$\times$1024 pixels.
The pixel size is 24$\mu$m$\times$24$\mu$m, or 1.6''$\times$1.6''
at the focal plane. It is designed to have a wide field view of
28'$\times$28'. The primary motivation of MITSuME
project is a multi-band photometry of gamma-ray bursts 
and their afterglows at very early phases, but the telescopes are also
actively used for multi-color optical monitoring of more than 30
blazars and other interesting Galactic or extragalactic sources.
These telescopes are automatically operated and respond to GRB
alerts and transient events like AGN flares.

MITSuME observed 3C 454.3 almost every day from Nov 22 to Dec 6, 2007,
so as to provide simultanous data with \agile{} and $Suzaku$. All
raw $g'$,$R_c$ and $I_c$ frames were corrected for dark, bias and
flat field by using IRAF ver 2.12 software. Instrumental magnitudes
were obtained via aperture photometery using \textsc{DAOPHOT} (Stetson 1987)
and \textsc{SExtractor} (Bertin \& Arnouts 1996). Calibration of the
optical source magnitude was conducted by differential photometry with
respect to the comparison stars sequence reported by Raiteri et
al. (1998) and Gonzalez-Perez et al. (2001). The fluxes are corrected
for the Galactic extinction corresponding to a reddening of $E(B-V)$
= 0.108 mag (Schlegel et al. 1998). The $R_{c}$-optical light curve between
November 30 and December 6 is showed in Fig. 1 (bottom panel, green circles).

\section{Mid-Infrared observations}

Given the high $\gamma$-ray activity detected by \agile{} from 3C
454.3, we also requested and obtained a Director's Discretionary Time for a
mid-Infrared follow-up by Spitzer (Werner et al. 2004).
The DDT was approved for 2 epochs for a total duration of 0.8 hours  of the
Infrared Spectrograph (IRS, Houck et al. 2004) providing short-low and long-low
observations of 3C 454.3 scheduled for December 13 (starting at MJD 54447.410)
and 15 (starting at MJD
54449.403). Both observations
provided us with a low resolution spectrum ($\Delta\lambda/\lambda \sim 80$) in the energy range $\sim 5-38$ $ \mu m$.
Data were acquired in the IRS standard staring mode: observations were
obtained at two positions along the slit to enable sky subtraction. Each ramp
duration was set to 14.68 s with a number of cycles equal to 5.
Each set of data was processed with the IRS Standard Pipeline {\it SMART}
developed at the Spitzer Science Center to produce calibrated data frames
(Basic Calibrated Data, BCD files). Moreover, the BCD files covering the same spectral range were coadded
and then sky-subtracted spectra were obtained.
The absolute flux calibration was estimated by using the
electron-to-Jy conversion polynomial given in the appropriate {\it Spitzer}
calibration file. In Fig. 5 we present the two spectra obtained on December 13
and 15. We performed a linear fit of the two, obtaining a flux equal to
$(1.59\pm 0.02)\times 10^{-10}$ $(1.38\pm 0.02)\times 10^{-10}$ \ergcmsec for December 13 and 15, respectively.




\section{Discussion}

\subsection{Timing analysis}
We investigated the emission of the blazar 3C 454.3
during a multifrequency campaign performed in the first half of December 2007. The source was found to be in
flaring state with an average $\gamma$-ray flux above 100 MeV of $\sim 250\times 10^{-8}$ \phcmsec, which is typical of its
high gamma-ray state (Vercellone et al. 2008, Anderhub et al.
2008, Vercellone et al. 2009). As in the case of the previous multifrequency
campaign (November 2007, Vercellone et al. 2009), the source was continuously monitored in $\gamma$-rays
as well as in the optical energy bands. In both energy bands the
source exhibited comparable flux variations of the order of $\sim 4$: this argues
for an EC model. 
Moreover, we deeply studied the optical-$\gamma-ray$ correlation by means of a Discrete
Correlation Function (DCF; Edelson \& Krolik 1988; Hufnagel \&
Bregman 1992) applied to the optical and gamma-ray light curves reported in
Fig. 1.  This analysis revealed $\lesssim$ 1 day delay of the
gamma-ray emission with respect to the optical one (see Fig. 6).
Indeed, the DCF maximum at a time lag $\tau=-1$ day corresponds to a centroid
$\tau=-0.56$ day, whose uncertainty can be estimated by means of
the Monte Carlo method known as ``flux redistribution/random
subset selection'' (Peterson et al. 1998, Raiteri et al. 2003). By
running 1000 simulations we found $\tau= -0.6^{+0.7}_{-0.5}$ day at 1 sigma
confidence level. We also performed the DCF reducing the data
binning down to 12 hr between December 5 and 16, keeping the 1-day binned
\gray light curve for the data before December 5 (MJD=54439.524). This shows a peak at -1 day
with centroid at -0.54 which is in agreement with the result obtained with the
1-day binned \gray light curve. In this case, the Monte Carlo method is not able to
provide a reliable estimate of the error on the time lag due to the larger
uncertainties on the \gray fluxes.  

The evidence of this time lag again suggests the dominance of
the EC model: such a delay is compatible with the typical blob
dimensions and the corresponding crossing time of the external
seed photons (Sokolov et al. 2004). We note that this evidence
agrees with what was found by Bonnig et al. (2009).

Particularly interesting is the source optical flare recorded by
WEBT on December 12 (Raiteri et al. 2008a). The source experienced
an exceptional variability in less than 3 hr. Raiteri et al. (2008a)
interpreted this event as a variation in the properties of the jet
emission. This unusual event clearly
required an intra-day analysis of the $\gamma$-ray data. This 
analysis depends on the source brightness and the instrumental sensitivity. Given
the $\gamma$-ray  flux level of 3C 454.3
reached between 2007-12-05 and 2007-12-16, we obtained a data binning not
smaller than 12 hr (Fig. 7). This analysis
showed an enhancement of more than a factor of 2 of the $\gamma$-ray
flux during the second half of 2007 December 12, that remarkably includes the
time of the optical event (see vertical lines in Fig. 7). The enhancement by a
factor of $\sim 2$ of the $\gamma$-ray flux was
comparable with the 1.1 mag optical brightening. This could support the evidence of a change in the
jet emission in the EC scenario.
The 12-hour \gray light curve could constrain a possible delay between
the \gray emission and the optical one within 12 hours, shorter than ever
observed before for this source. 

%


\subsection{Spectral Modelling}
As described in the previous section, the December 2007 multifrequency campaign was
characterized by ToO carried out in mid-Infrared ({\it Spitzer}), soft X-ray
(\emph{Suzaku}, \emph{Swift}), hard X-ray (\emph{Suzaku}) and 
radio-to-optical and \gray monitoring.
These observations allowed us to obtain the SED of this blazar with a wide
multi-frequency coverage for three different epochs: December 5, 13, 15. At these dates the SED in X-rays shows a
\textit{softening} towards lower frequencies that can be due to
two causes: 1) a contribution from bulk comptonization by
cold electrons in the jet (Celotti, Ghisellini \& Fabian 2007),
2) the emergence of the SSC contribution in soft X-rays
from the more energetic EC component due to the disk and the BLR. The
mid-Infrared \textit{Spitzer} data and optical
data available in December 13 and 15 (which well define the synchrotron peak),
combined with the resolved X-ray spectrum and the gamma-ray data constrain the model parameters, arguing for the
latter cause; the SSC emergence is a natural and inevitable
consequence of the simultaneous modelling of the broad-band SED.
Nevertheless, some contribution from bulk comptonization cannot be
ruled out.

We first considered the state of December 13 and 15 in
which we have radio, mid-Infrared, optical, X and $\gamma$-ray
simultaneous data. In these epochs \source{}  was in a different
state with respect to the one analyzed in November: optical and UV
fluxes appeared lower by a factor 2-3, suggesting the synchrotron
bump peaking at a frequency 5-10 times lower than the one
observed in November, as confirmed also from the mid-Infrared
data. On the other hand, the soft X-ray data were only a little
bit lower than in November. Despite the softer synchrotron bump,
\gray data showed in the SED the \textit{persistence} of a
hard peak at $\simeq 1$ GeV, similar to the higher states
observed by \agile{} in July 2007 and November 2007 (Anderhub et al. 2009,
Vercellone et al. 2009). In fact, the
December $\gamma$-ray spectrum (characterized by a photon index of
$\sim 1.78$) is consistent with those obtained  during the 2
previous \agile{} observations.

We attempted to fit the SEDs with a one-zone SSC
model, adding the contribution of external seed photons coming
from an accretion disk and a BLR (Raiteri et
al. 2007). With this model, we succeeded to fit the
synchrotron peak as well as the X-ray data assuming parameters
similar to the November ones, but a \textit{lower}
$\gamma_b\simeq 350$ was required to account for the softness of the synchrotron bump: with this $\gamma_b$
the EC from a standard BLR peaks at $h\nu\simeq
h\nu_{soft}\Gamma\gamma_b^2\delta/(1+z)\sim10^{8}\,$eV.  This is in contrast
with the observed hardness of the \gray spectrum up to 1 GeV
($h\nu_{soft}\simeq 10$eV is the typical energy of the external
source as seen by the observer). We note that the EC by the disk
can account for the rising hard X-ray portion
of the SED, which did not show clear variability. Nevertheless, we
note that both the disk and BLR components cannot account for the
hardness of the \gray spectrum. Thus, we consider a
further external source of seed photons. 

A possibile candidate for this source is the hot extended
corona that must be consistently produced in steady accretion/ejection
flows as shown by MHD numerical simulations (Tzeferacos et al. 2009).
Hence, we considered a one-zone SSC model plus the contribution by
external seed photons coming from the accretion disk, the BLR and the hot
corona. We adopted a spherical blob with radius $R=2.2\times
10^{16}$cm and a broken power law for the electron energy density
in the blob,
\begin{equation}
n_{e}(\gamma)=\frac{K\gamma_{b}^{-1}}{(\gamma
/\gamma_{b})^{a_{l}}+(\gamma /\gamma_{b})^{a_{h}}}\,
\end{equation}
where $\gamma$ is the electron Lorentz factor assumed to vary
between $10<\gamma<10^{4}$, while $a_{l}=2.3$ and $a_{h}=4.2$ are
the pre-- and post--break electron distribution spectral indices,
respectively. We assumed that the blob contained a random magnetic
field $B=2$ Gauss and that it moved with bulk Lorentz Factor
$\Gamma=18$ at an angle $\Theta_{0}=1^{\circ}$
($\delta\simeq33$) with respect to the line of sight. The density
parameter into the blob is $K=52$ cm$^{-3}$.

The bolometric luminosity of the accretion disk is $L_{\rm
d}=3\,10^{46}$ erg $\, $s$^{-1}$, and it is assumed to lie at
0.01\,pc from the blob; we assumed a BLR distant 1.5 pc,
reprocessing a 10$\%$ of the irradiating continuum. We assumed for the disk a black-body spectrum  peaking in UV (see
Tavecchio and Ghisellini 2008, Raiteri et al. 2008b). Finally, we
added the hot corona photons surrounding the jet as a black body
spectrum of $T=10^6\, K$ and $L_h=10^{45}$ erg $\,
$s$^{-1}$, and distant 0.5 pc from the blob. The SEDs of both December 13 and
15 could be fitted  with almost the same parameters (see red and blue solid
lines in Fig. 8). The high
energy portion of the electron density becomes softer in December
15 as the same electrons should be accelerated with less
efficiency than in December 13.

Remarkably, the lower $\gamma_b$
required in the epochs considered here, makes the
BLR a too soft contributor at GeV energies, while the contribution of the hot
corona succeeded to account for the persistence of the hard \gray spectra
measured by \agile{}.

On December 5, the low energy peak of the SED is less constrained
with respect to the December 13 and 15 ones due to the lack of the
mid-Infrared data. On the other hand the {\it Suzaku} X-ray data (green points in
Fig. 8) better constrain the rise of the IC emission. We fitted this SED with
almost the same model assumed for the other two epochs, but the higher optical flux and the
lower \gray flux detected with respect to December 13 required a higher
magnetic field and a lower $\gamma_{b}$ (see Table 2).\\

Given the different $\gamma$-ray state of the source analyzed in the November and
December campaigns, we compared the particle
injection luminosity, $L_{\rm inj}$\, measured during the two multiwavelength campaigns.
This is expressed by means of the following formula:
\begin{equation}
L_{\rm inj} = \pi\,R^{2}\,\Gamma^{2}\,c\,\int[d\gamma\,\,m_{\rm e}\,c^{2}\gamma\,n(\gamma)].
\end{equation}
We found the particle injection luminosity of December to be $6 \times 10^{43} \rm erg \rm
s^{-1}$, a factor of 5 lower than the November one. This difference is
due to both the lower $\gamma_{b}$ and $\gamma_{min}$ values needed to reproduce the SED in the
states of December.

\section{Conclusions}

We reported in this paper the main results of a multifrequency campaign on the
blazar 3C 454.3 performed in December 2007. The source was simultaneously
observed in mid-Infrared, optical, X-ray and \gray energy bands, which
provided us with a wide dataset aimed to study the correlation between the
emission properties at lower and higher frequencies.
We summarize below the major results.

\begin{itemize}
\item The \gray emission from 3C 454.3 shows variations on a daily time scale.
\item The simultaneous monitoring of the source in the optical and \gray
  energy bands allowed us to determine a possible $\lesssim$ 1 day delay of the \gray
  emission with respect to the optical one.  
\item The extraordinary optical activity (lasting less than 3 hours), 
  observed on December 12 has a counterpart in the \gray data. A possible
  delay between the \gray emission and the optical one is constrained within 12
  hours.
\item We found that a leptonic model with an External Compton on seed photons
  from disk and BLR does not succeed to account for both the ``hardeness'' of the \gray
  spectrum and the ``softness'' of the Synchrotron emission, requiring  an additional component. We argued that a
  possible candidate for it is  the hot Corona ($T \sim 10^{6}$ K)
  surrounding the disk. 
\end{itemize}
\acknowledgements 

{\agile{} is a mission of the Italian Space Agency, with
co-participation of INAF (Istituto Nazionale di Astrofisica) and
INFN (Istituto Nazionale di Fisica Nucleare). This work was
partially supported by ASI grants I/R/045/04, I/089/06/0, I/011/07/0
and by the Italian Ministry of University and Research (PRIN
2005025417). INAF personnel at ASDC are under ASI contract
I/024/05/1.
This work is partly based on data taken and assembled by the WEBT
collaboration and stored in the WEBT 
archive at the Osservatorio Astronomico di Torino-INAF ({\tt http://www.oato.inaf.it/blazars/webt/}).
This work is based in part on observations made with the Spitzer Space
Telescope, which is operated by the Jet Propulsion Laboratory, California
Institute of Technology under a contract with NASA. The IRS was a
collaborative venture between Cornell University and Ball Aerospace
Corporation funded by NASA through the Jet Propulsion Laboratory and Ames
Research Center. SMART was developed by the IRS Team at Cornell University and is available through the Spitzer Science Center at Caltech.
}



\nocite{*}

\clearpage
\begin{figure}
\centering
\epsscale{0.5}
\subfigure{\includegraphics[angle=0,scale=0.5]{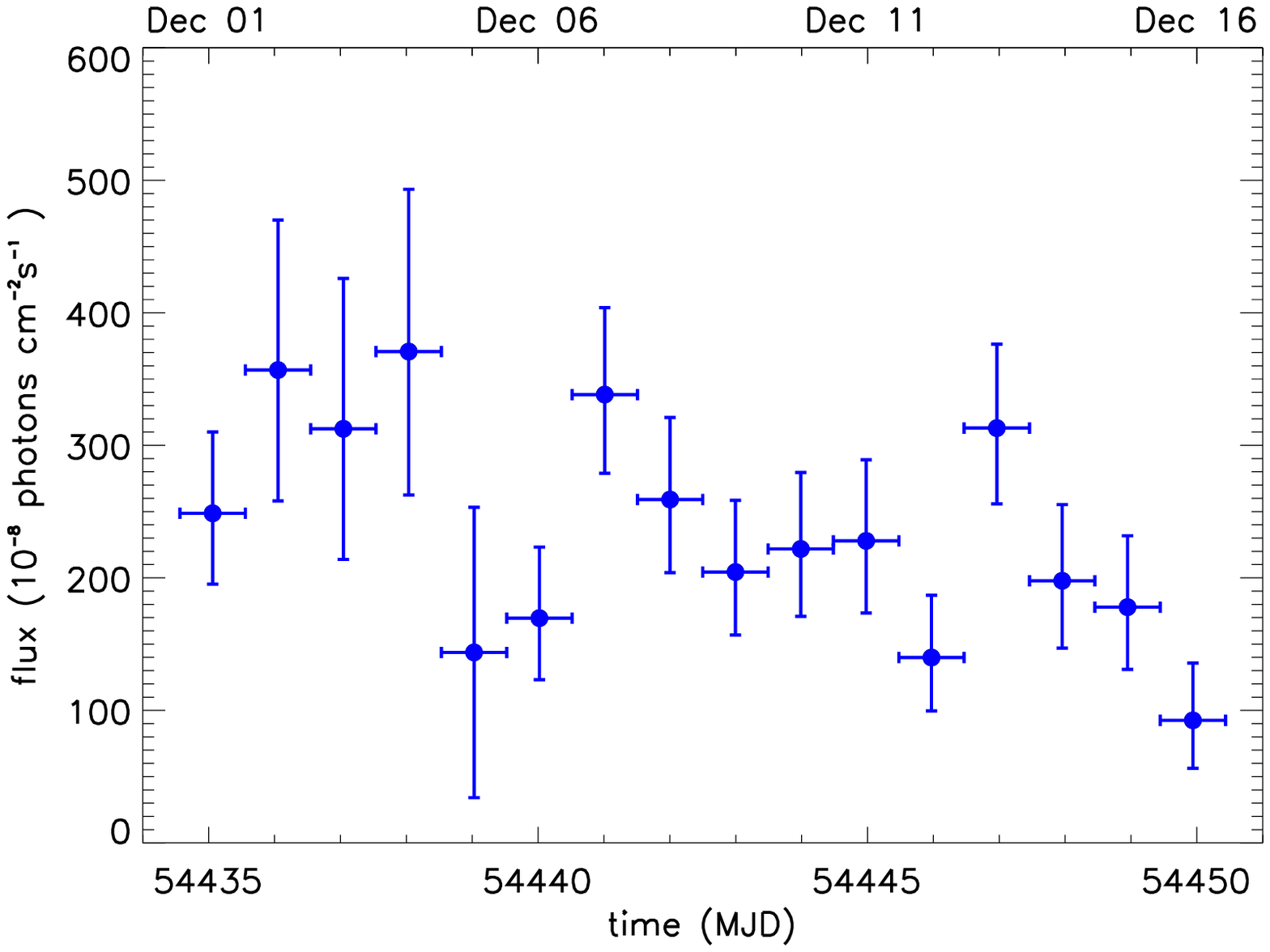}}
\hspace{10cm}
\subfigure{\includegraphics[angle=0,scale=0.5]{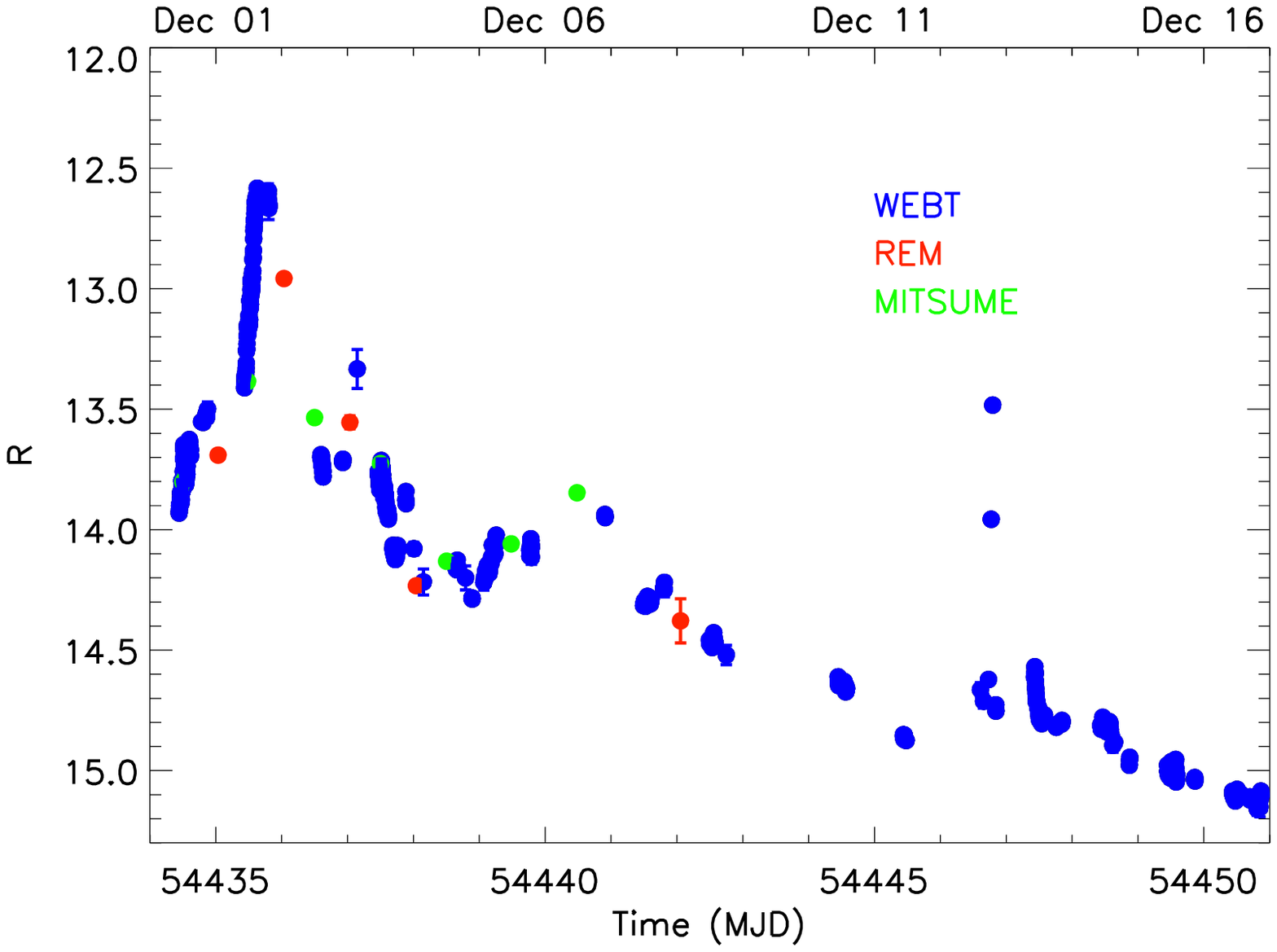}}

\caption{Top panel: \agile-GRID $\gamma$-ray light-curve of 3C 454.3 at 1-day resolution
  for E$>$100 MeV in units of 10$^{-8}$ \phcmsec during the period 2007
  November 30--December 16; 
bottom panel:  $R$-band light curve obtained by WEBT (blue circles), REM (red
    circles), \MITSUME{} (green circles) between November 30 and December 16.}
\end{figure}

\begin{figure}
\centering
\includegraphics[angle=0, scale=0.9]{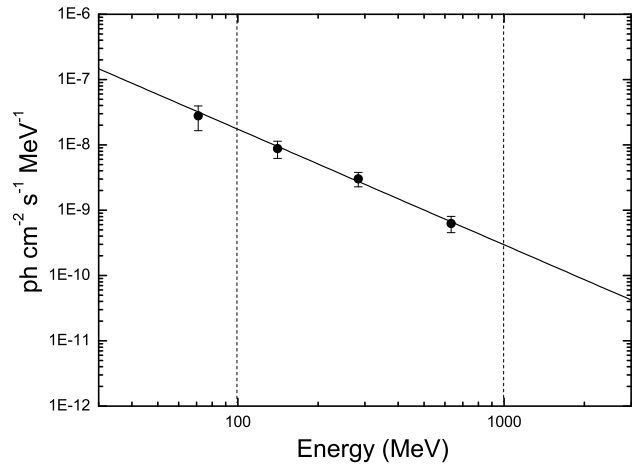}
\caption{Gamma-ray photon spectrum of 3C 454.3 during the observation period
5 - 16  December 2007. Only the energy bins above 100 MeV were taken into account in the
spectral fitting.  The solid line corresponds to a power law function with photon index $\Gamma=1.78\pm 0.14$.}
\end{figure}

\begin{figure}[hbt]
\centering
\epsscale{0.6}
\rotatebox{270}
{\plotone{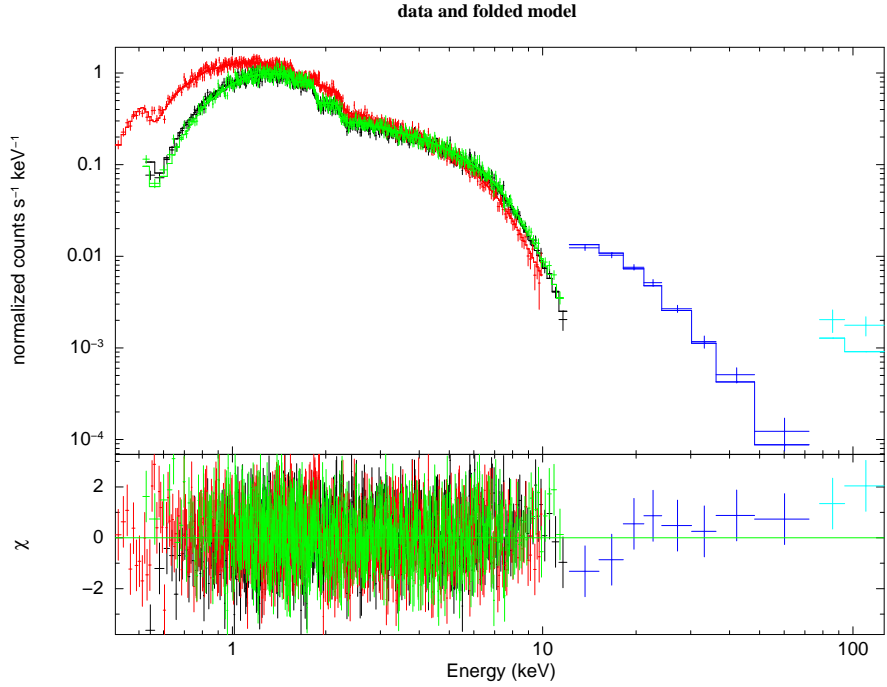}}

\caption{\suzaku{} broad-band spectrum of 3C454.3 for the observation carried
  out in 2007 December 5: black, red and green points for Xis0, Xis1, Xis3,
  respectively; blue points for PIN; cyan points for GSO. In the folded model,
a single power law over the whole energy range is assumed. See the text for details.}
\end{figure}

\begin{figure}[hbt]
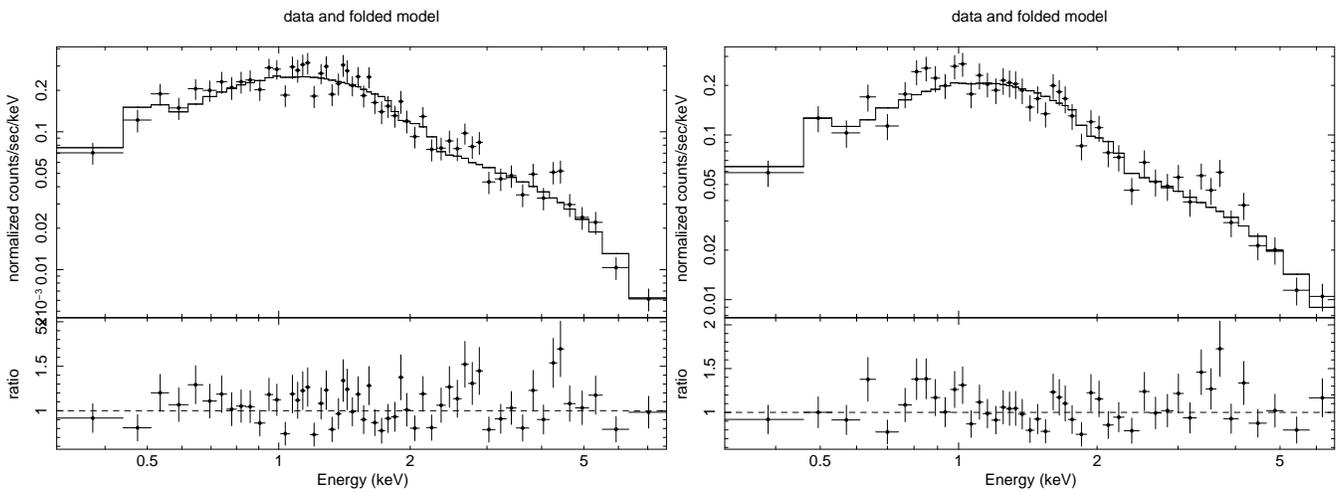

\subfigure{\includegraphics[angle=270,scale=0.38]{f4_1.eps}}
\subfigure{\includegraphics[angle=270,scale=0.38]{f4_2.eps}}
\caption{\swi{} data and folded model of 3C 454.3 for the observation carried
  out in 2007 December 13 (left panel) and 15 (right panel).}
\end{figure}

\begin{figure}[hbt]
\centering
\includegraphics[angle=0,scale=0.6]{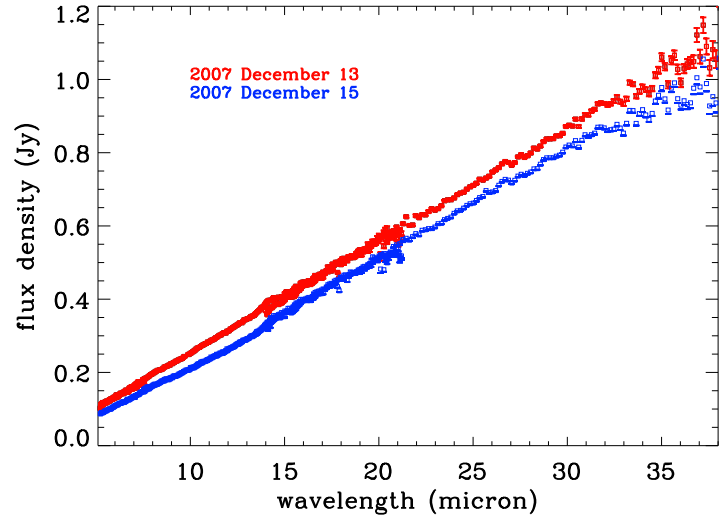}

\caption{{\it Spitzer} spectra of 3C 454.3 for the observation carried out
  in 2007 December 13 (red points) and 15 (blue points).}
\end{figure}


\begin{figure}[hbt]
\centering
\includegraphics[angle=0., scale=0.5]{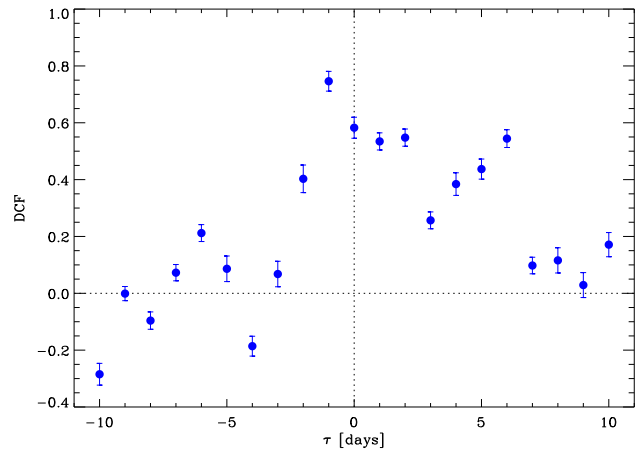}
\caption{3C 454.3 Discrete Correlation Function (DCF) between the \gray and optical ($R$-band) magnitudes.}
\end{figure}

\clearpage

\begin{figure}[hbt]
\centering
\subfigure{\includegraphics[angle=0, scale=0.5]{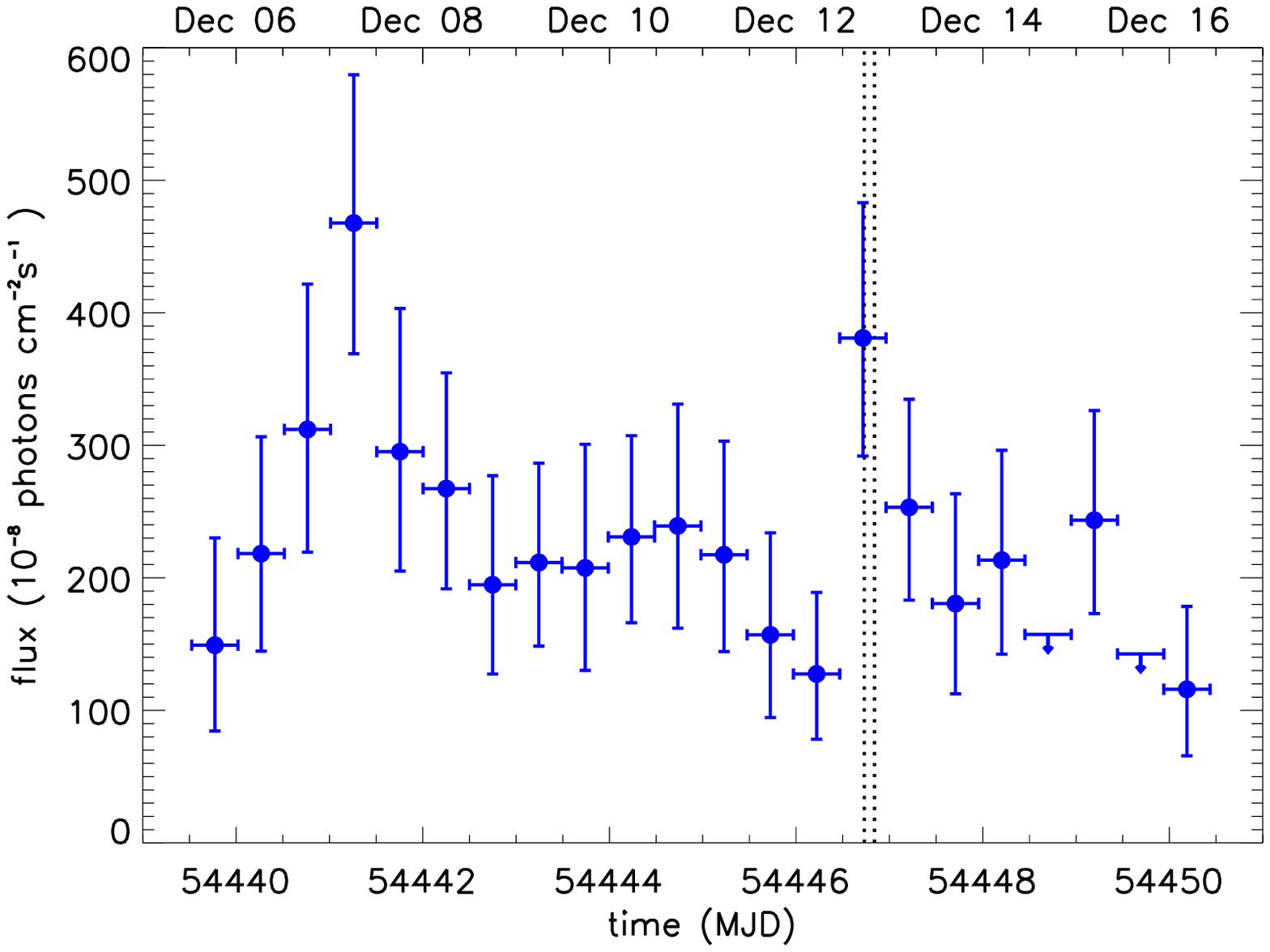}}
\subfigure{\includegraphics[angle=0,scale=0.5]{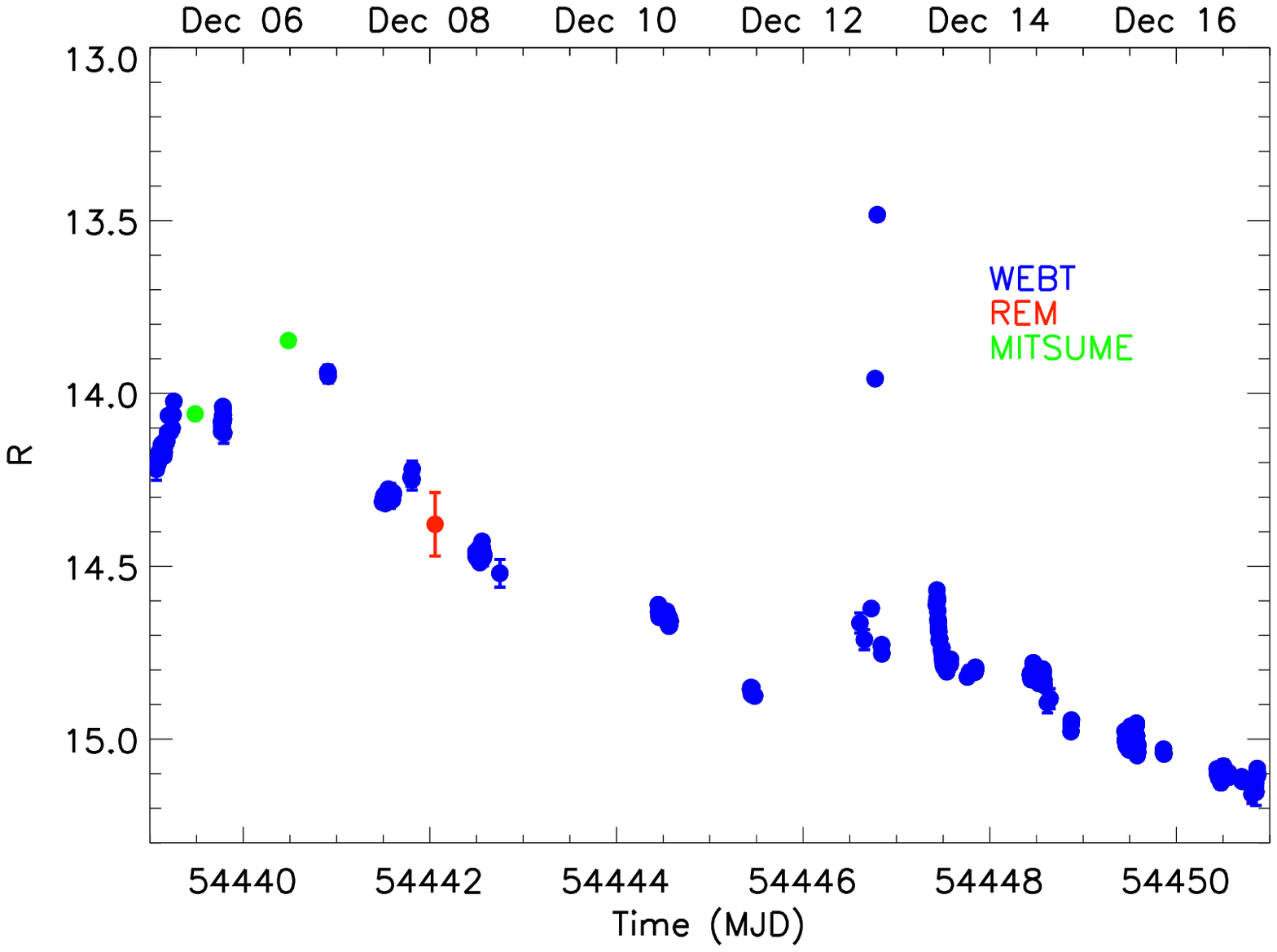}}
\caption{Top panel: \gray 12-hr light curve above 100 MeV during the period
  between between  2007-12-05 and
  2007-12-16 when the source was at 30 degrees in the \agile{} FoV. The vertical
  lines mark the time ($<3$ hr) of the exceptional optical event of 2007
  December 12; bottom panel:  $R$-band light curve obtained by WEBT (blue circles), REM (red
    circles), \MITSUME{} (green circles) during the period reported above.}
\end{figure}

\begin{figure}[hbt]
\begin{center}
\includegraphics[angle=0, width=10cm, height=7cm]{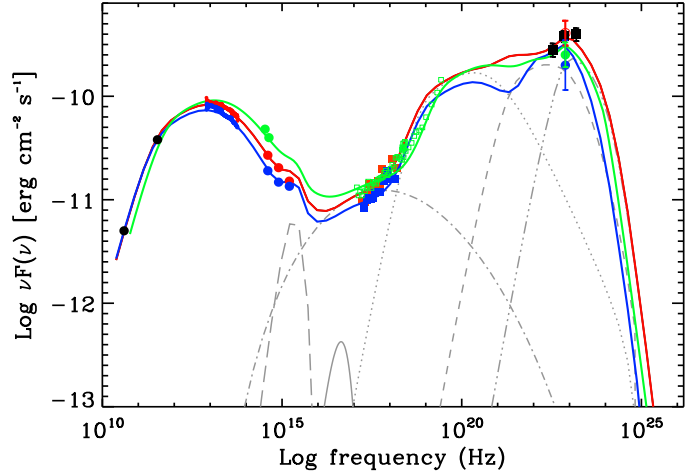}
\caption{3C 454.3 SEDs for 2007 December 5, 13 and 15 (green, red and blue solid
  lines, respectively). The $\gamma$-ray spectrum for $E > 100$ MeV (black squares),
  exctracted from data acquired between December 5--16 and the radio points
  (black circles) from Raiteri et al. (2008a) are also reported. The
  gray lines represent the contribution of the disk (long dashes), corona
  (solid), SSC (dot-dashed), EC 
  disk (dotted), EC BLR (dashed), EC Corona (dash dot dot)  to the December 13 model. }
\end{center}
\end{figure}

\clearpage
\begin{table*}

\caption{Results of XRT observations of 3C 454.3. Power law model with $N_{\rm
H}$ free to vary. }
\begin{tabular}{ccccccc}
\hline
\multicolumn{1}{c}{Observation} &
\multicolumn{1}{c}{N$_{H}$} &
\multicolumn{1}{c}{Flux 0.3-10 keV} &
\multicolumn{1}{c}{Flux 2-10 keV} &
\multicolumn{1}{c}{Spectral slope} &
\multicolumn{1}{c}{$ \chi^{2}_{r}$/d.o.f.} \\
\multicolumn{1}{c}{date} &
\multicolumn{1}{c}{10$^{22}$ cm$^{-2}$} &
\multicolumn{1}{c}{erg cm$^{-2}$ s$^{-1}$} &
\multicolumn{1}{c}{erg cm$^{-2}$ s$^{-1}$} &
\multicolumn{1}{c}{$\Gamma$}&
\multicolumn{1}{c}{} \\
\hline
13-Dec-2007 & $0.13 \pm 0.03$ & $(4.38\pm 0.25)\times 10^{-11}$ & $(3.04 \pm 0.24)\times 10^{-11}$ & $1.74 \pm 0.10$ &1.28/54\\
15-Dec-2007 & $0.14 \pm 0.03$ & $(3.60\pm 0.22)\times 10^{-11}$ & $(2.49 \pm 0.22)\times 10^{-11}$ & $1.76 \pm 0.12$ &1.14/44\\
\hline
\end{tabular}
\label{tab.spectral_results}
\end{table*}

\begin{table*}
\caption{Model parameters for the December states of 3C 454.3}
\begin{tabular}{ccccccccc}
\hline
\multicolumn{1}{c}{Observation Date} &
\multicolumn{1}{c}{$\Gamma$} &
\multicolumn{1}{c}{B (Gauss)} &
\multicolumn{1}{c}{R (cm)} &
\multicolumn{1}{c}{K (cm$^{-3}$)} &
\multicolumn{1}{c}{$\gamma_b$} &
\multicolumn{1}{c}{$\gamma_{min}$} &
\multicolumn{1}{c}{$a_l$} &
\multicolumn{1}{c}{$a_h$} \\
\hline
  5-Dec-2007 & 18 & 2.5   & $2.2\times10^{16}$ & 50 & $3\times10^2$ & 30 & 2.3&4.2\\
  13-Dec-2007 &  18 & 2   & $2.2\times10^{16}$ & 52 & $3.5\times10^2$ & 38 &2.3&4.2\\
  15-Dec-2007 &  18 & 2   & $2.2\times10^{16}$ & 52 & $3.2\times10^2$ & 35 &2.3&4.2\\
  \hline
\end{tabular}
\end{table*}
\normalsize

\end{document}